
\def\sqr#1#2{{\vcenter{\hrule height.#2pt
             \hbox{\vrule width.#2pt height#1pt \kern#1pt
             \vrule width.#2pt}
             \hrule height.#2pt}}}

\baselineskip=15pt

\def\pmb#1{\setbox0=\hbox{#1}\kern-.025em
    \copy0\kern-\wd0\kern.05em\kern-.025em\raise.029em\box0}

\def\a{\alpha}      \def\l{\lambda}   \def\L{\Lambda} 
\def\b{\beta}       \def\m{\mu}            
\def\g{\gamma}      \def\G{\Gamma}    \def\n{\nu}        
\def\d{\delta}                 
                      \def\r{\rho}        
\def\ve{\varepsilon}                  \def\s{\sigma}  \def\S{\Sigma}
                         
        \def\vphi{\varphi}
                   \def\o{\omega}  \def\O{\Omega}
\def\k{\kappa}

\def\cL{{\cal L}}

\def\fr#1 #2{\hbox{${#1\over #2}$}}        
\def\leaderfill{\leaders\hbox to 1em{\hss.\hss}\hfill}

\def\section#1{                            
\vskip.6cm\goodbreak                       
\noindent{\bf \uppercase{#1}}
\nobreak\vskip.4cm\nobreak  }

\def\subsection#1{
  \vskip.4cm\goodbreak
  \noindent{\bf #1} 
  \vskip.3cm\nobreak}

\def\subsub#1{\par\vskip4pt {\bf #1} }

\def\ver#1{\left\vert\vbox to #1mm{}\right.}


\font\eightrm=cmr8                          
\font\tfont=cmbx12                          

\def\preprint#1{\hskip10cm #1 \par}
\def\date#1{\hskip10cm #1 \vskip2cm}
\def\title#1{ \centerline{\tfont{#1}} }
\def\titlef#1{ \vskip.2cm                   
      \centerline{ \tfont{#1}
                   \hskip-5pt ${ \phantom{\ver{3}} }^\star$   }
      \vfootnote{$^\star$}{\eightrm Work supported in part 
      by the Serbian Research Foundation, Yugoslavia.} }
\def\author#1{ \vskip.8cm \centerline{#1} }
\def\institution#1{ \centerline{\it #1} }

\def\abstract#1{ \vskip3cm \centerline{\bf Abstract}  
                 \vskip.3cm {#1} \vfill\eject } 

\magnification=1200

\def\Tr{\,{\rm Tr}\,}

\def\tg{{\tilde g}} 
\def\pa{\partial}

\null

\preprint{IF/14--1997}
\date{September, 1997}
\title{On the Equivalence between 2D Induced Gravity}
\titlef{and a WZNW System} 
\author{M. Blagojevi\'c, D. Popovi\'c and B. Sazdovi\'c}
\institution{Institute of Physics, P.O.B. 57, 11001 Belgrade, Yugoslavia}

\abstract{A general method of constructing canonical gauge invariant
actions is used to establish the equivalence between 2D induced gravity
and a WZNW system, defined by a difference of two simple WZNW actions
for the $SL(2,R)$ group. The diffeomorfism invariance of the induced
gravity is generated by the $SL(2,R)$ Kac--Moody structure of the WZNW
system.}       

\subsection{1. Introduction} 

A dynamical structure of quantum gravity in 2D, which is induced by
string theory, is of great importance for understanding string dynamics
[1]. The induced gravity action in the light--cone gauge possesses a
hidden chiral $SL(2,R)$ symetry [2,3], while in the conformal gauge
it becomes the Liouville theory [4,5]. Having in mind the dynamical
importance of the $SL(2,R)$ symmetry, it is natural to try to
understand the way in which this symmetry is associated to the induced
gravity, and, thereby, to the Liouville theory.  

In this paper we shall show that the induced gravity action,
$$
S(\phi,g_{\m\n}) =\int d^2\xi\sqrt{-g}\,\Bigl[
  {\fr 1 2}g^{\m\n}\pa_\m\phi\pa_\n\phi +
  {\fr 1 2}\a\phi R +M^2\bigl(e^{2\phi/\a}-1\bigr)\Bigr]\, ,  \eqno(1)
$$
is {\it gauge equivalent\/} to the $SL(2,R)$ Wess--Zumino--Novikov--Witten 
(WZNW) system,  
$$
S(g_1,g_2)=S_L(g_1)-S_R(g_2) \qquad g_1,g_2\in SL(2,R) \, ,    \eqno(2)
$$
defined by a difference of two, left and right, WZNW actions (L--R).
In the process of establishing this equivalence, the connection between
the $SL(2,R)$ Kac--Moody (KM) structure of the WZNW system and the
diffeomorfism invariance of the induced gravity will become much more
clear.

Our result generalizes Polyakov's work, who found the connection between
the WZNW action for $SL(2,R)$, and the induced gravity in {\it the
light--cone gauge\/} [6]. Similar results in the light--cone gauge have
been obtained in Refs.[7,8], using the methods of conformal field
theory and coadjoint orbits of the Virasoro group, respectively.
The same problem was also discussed in {\it the conformal gauge\/}, in
Ref. [9], where the related connection is used to obtain the general
solution of the Liouville theory from that of the WZNW model. Due to
the presence of {\it two\/} simple WZNW actions in (2), we are able to
demonstrate the equivalence of (1) and (2) in a covariant way, fully
respecting {\it the diffeomorfism invariance\/} of the induced gravity.    

We shall use the general method of constructing gauge invariant actions
based on the Hamiltonian formalism [10]. The method uses the fact
that the Lagrangian equations of motions are equivalent to the
Hamiltonian equations derived from the action
$$
S(q,\pi,u)=\int d\xi (\pi_i \dot q^i -H_0-u^m G_m) \, ,   \eqno(3a)
$$
where $G_m$ are primary constraints, and $H_0$ is the canonical
Hamiltonian. If $G_m$ are first class constraints, satisfying the
Poisson bracket algebra
$$\eqalign{
&\{ G_m,G_n\}=U_{mn}{^r}G_r \, ,\cr
&\{ G_m,H_0\}=V_m{^r}G_r \, , }                           \eqno(3b)
$$
than the action $S(q,p,u)$ is invariant under the following gauge
transformations: 
$$\eqalign{
&\d F=\ve^m\{F,G_m\} \, , \qquad F=F(q,\pi) \cr
&\d u^m= \dot\ve^m + u^r\ve^s U_{sr}{^m}+\ve^rV_r{^m} \, .}\eqno(4)
$$

Using this idea we shall start with the action (2), make a convenient
{\it gauge extension\/} of this theory by using $(3a)$, and show that
the resulting formulation of (2) reduces to (1) after a convenient
{\it gauge fixing\/}. 

\subsection{2. The WZNW model for $SL(2,R)$} 

We shall begin by recalling some facts about the WZNW model, described
by the action  
$$
S_L(g) = S_0(g) + n\G(g)
       = {n\over 8\pi}\int_{\S}({}^*v,v) 
         +{n\over 24\pi}\int_B (v,[v,v])  \, .               \eqno(5)
$$
The first term is the action of the non--linear $\s$--model for a
field $g$, defined over a two dimensional spacetime $\S$ and taking
values in a semisimple group $G$,
$v=g^{-1}dg$ is the Maurer--Cartan one--form, ${}^*v$ is the 
dual of $v$, and $(X,Y)={\fr 1 2}\Tr(XY)$ is the Cartan--Killing
bilinear form (the trace operation is taken in the adjoint
representation of $G$). The second term is the topological Wess--Zumino
term,  where $B$ is a three dimensional manifold with boundary $\S$.

Now, we turn our attention to the case of $G=SL(2,R)$. The generators
of $SL(2,R)$ are taken as $t_a=(t_-,t_0,t_+)=(\s_-,\s_3/2,\s_+)$, where
$\s_i$ are the Pauli matrices. They define the structure constants
$f_{ab}{^c}$  ($f_{-0}{^-}=1,\, f_{+0}{^+}=-1,\, f_{+-}{^0}=2$), and
the Cartan metric $\g_{ab}=(t_a,t_b)$ (with nonvaishing
compoents $\g_{-+}=\g_{+-}=2,\, \g_{00}=1$). In $SL(2,R)$, any group
element $g$ admits the Gauss decomposition, 
$g=e^{xt_+}e^{\vphi t_0}e^{yt_-}$, where $q^\a=(x,\vphi,y)$ are
group coordinates in a neighbourhood of the identity, 
$$
g =\pmatrix{e^{\vphi/2}+xye^{-\vphi/2} &xe^{-\vphi/2} \cr
            ye^{-\vphi/2}              &e^{-\vphi/2} \cr}\, ,
$$
and, also, $q$'s are the usual fields on $\S$, $q=q(\xi)$.
With the above expression for $g$, the WZNW action for $SL(2,R)$
takes the simple local form 
$$
S_L(q)= \k \int_{\S}d^2\xi\, \bigl[ {\fr 1 2} \eta^{\m\n}
                                 \partial_\m\vphi\partial_\n\vphi
 +2(\eta^{\m\n}-\ve^{\m\n})\partial_\m x\partial_\n y e^{-\vphi} \bigr]\, , 
                                                              \eqno(6)
$$
where $\k\equiv n/8\pi$.

The dynamical structure of the theory $(6)$ is characterized by the
existence of two sets of currents. In the phase space with coordinates
$(q^\a,\pi_\a)$, where $\pi_\a=(\pi_x,\pi_\vphi,\pi_y)$ are momenta
canonically conjugate to $q^\a$, the left--handed currents $J_{(-)a}$
are given as [11]    
$$\eqalign{
&J_{(-)+}=\pi_x \, ,\cr
&J_{(-)0}=x\pi_x+(\pi_\vphi -\k\vphi') \, ,\cr
&J_{(-)-}=-x^2\pi_x -2x(\pi_\vphi -\k\vphi') -4\k x' +\pi_y e^\vphi \, ,}
                                                             \eqno(7a)
$$
while the right--handed ones, $J_{(+)a}$, are
$$\eqalign{
&J_{(+)+}= y^2\pi_y+2y(\pi_\vphi+\k\vphi')-4\k y'-\pi_x e^{\vphi}\, ,\cr
&J_{(+)0}=-y\pi_y-(\pi_\vphi+\k\vphi') \, ,\cr
&J_{(+)-}=-\pi_y\, .}                                        \eqno(7b)
$$
Using the basic Poisson brackets 
$\{q^\a(\s_1),\pi_\b(\s_2)\}=\d^\a_\b\d(\s_1-\s_2)$, one finds that the 
currents $J_{(\mp)a}$  satisfy two independent KM algebras:
$$
\{J_{(\mp)a},J_{(\mp)b}\}=f_{ab}{^c}J_{(\mp)c}\d \mp 2\k\g_{ab}\d' \, ,
                                                             \eqno(8)
$$
where $\d=\d(\s_1-\s_2)$, $\d'=\pa_{\s_1}\d$, and 
$\{J_{(-)a},J_{(+)b}\}=0$.

In a similar way one can define $S_R(g)=S_0(g)-n\G(g)$, whose local
coodinate expression $S_R(q)$ is obtained from $(6)$ by changing the
sign of the $\ve^{\m\n}$--term. 

\subsection{3. Covariant extension of the WZNW model for $SL(2,R)$} 

Now, as a preparation for the main problem of proving the gauge
equivalence between the induced gravity and the WZNW system (2), we
shall first study the problem of the covariant extension (with
respect to diffeomorfisms) of the WZNZ models (6) and (2) [12]. 
\subsub{(A)} By using explicite expressions for the KM currents
(7) associated to the simple WZNW model (6), we can construct the related
$SL(2,R)$ invariant expressions, 
$$\eqalign{
&T_-={1\over 4\k}\g^{ab}J_{(-)a}J_{(-)b}=
       {1\over 4\k}\bigl[\pi_x\pi_y e^\vphi
        -4\k x'\pi_x +(\pi_\vphi -\k\vphi')^2 \bigr]\, ,\cr
&T_+=-{1\over 4\k}\g^{ab}J_{(+)a}J_{(+)b}=
       -{1\over 4\k}\bigl[\pi_x\pi_y e^\vphi
       +4\k y'\pi_y +(\pi_\vphi +\k\vphi')^2 \bigr]\, ,}     \eqno(9)
$$
representing components of the energy--momentum tensor. From the KM
algebra of currents we obtain two independent Virasoro algebras for
$T_-$ and $T_+$: 
$$
\{T_\mp(\s_1),T_\mp(\s_2)\}=-[T_\mp(\s_1)+T_\mp(\s_2)]\d'\, .
                                                             \eqno(10)
$$

Now, we wish to construct the canonical action $(3a)$ for a theory in
which $H_0=0$, and $G_m=(T_-,T_+)$ are first class constraints:
$$
\cL_L(q,\pi,h)=\pi_\a\dot q^\a -h^- T_--h^+T_+ \, .          \eqno(11a)  
$$
After eliminating the momentum variables with the help of the equations
of motion, and introducing a change of variables
$(h^-,h^+)\to\tg^{\m\n}$, 
$$
\tg^{00}={2\over h^--h^+}\, ,\qquad \tg^{01}={h^- +h^+\over h^--h^+}\, , 
\qquad \tg^{11}={2h^-h^+\over h^- -h^+} \, ,                 \eqno(12a)
$$
with $\det\tg^{\m\n}=-1$, the Lagrangian becomes 
$$
\cL_L(q,h)=\k\bigl[ {\fr 1 2}\tg^{\m\n}\pa_\m\vphi\pa_\n\vphi
    +2(\tg^{\m\n}-\ve^{\m\n})
         \pa_\m x\pa_\n y e^{-\vphi}\bigr] \, .              \eqno(11b)
$$
If we make the identification $\tg^{\m\n}=\sqrt{-g}g^{\m\n}$, the above
expression is seen to represent the covariant generalization of the
WZNW theory; for $h^\mp=\pm 1$ it reduces to $(6)$.

The transformation properties of $\tg^{\m\n}$ are consistent with this
interpretation. Indeed, the transformation rules for the
multipliers $h^\mp$ are
$$
\d h^\mp=\pa_0\ve^\mp+h^\mp\partial_1\ve^\mp-\ve^\mp\partial_1 h^\mp\, .
                                                              \eqno(12b)
$$
Then, after introducing new parameters, $\ve^\mp=\ve^1-\ve^0 h^\mp$,
one finds 
$$
\d\tg^{\m\n}=\tg^{\m\r}\partial_\r\ve^\n+\tg^{\n\r}\partial_\r\ve^\m
        -\partial_\r\bigl(\ve^\r\tg^{\m\n}\bigr) \, ,          \eqno(12c)
$$
which is the diffeomorfism transformation of a metric density.

One should observe that the Virasoro generators $T_\mp$ are constructed
in terms of the KM currents (7). This construction yields a simple 
explanation of how the diffeomorfisms can be obtained out of the KM
structure of the WZNW model.  
\subsub{(B)} The next step in our descussion is to combine two simple
WZNW actions, as in Eq.(2), and construct the related covariant
extension. The KM currents  related to the L sector,
$J^{(1)}_{(\mp)a}$, are the same as in $(7a,b)$, while those related to
the R sector, $J^{(2)}_{(\mp)a}$, are of oposite chiralities:
$$
J^{(1)}_{(\mp)a}(q_1,\pi_1)=J_{(\mp)a}(q_1,\pi_1) \, ,\qquad
J^{(2)}_{(\mp)a}(q_2,\pi_2)=J_{(\pm)a}(q_2,\pi_2) \, ,       \eqno(13a)
$$
We also define the energy--momentum tensors of the L and R sectors
as 
$$
T^{(1)}_\mp(q_1,\pi_1)=T_\mp(q_1,\pi_1) \, ,\qquad
T^{(2)}_\mp(q_2,\pi_2)=-T_\pm(q_2,\pi_2) \, .                \eqno(13b)
$$
It is easy to check that the components of the complete
energy--momentum, 
$$
T_\mp=T^{(1)}_\mp +T^{(2)}_\mp \, ,                           \eqno(14)
$$
satisfy two independent Virasoro algebras (10).

Now, the canonical action in which $H_0=0$ and $G_m=(T_-,T_+)$ has
the form
$$
\cL(q_i,\pi_i,h)=\pi_{1\a}\dot q_1^\a+\pi_{2\a}\dot q_2^\a 
                                       -h^-T_- -h^+T_+ \, ,   \eqno(15a)
$$
As in the case (A), the elimination of momenta $\pi_{1\a}$ and
$\pi_{2\a}$ leads to the result
$$
\cL(q_1,q_2,h)=\cL_L(q_1,h)-\cL_R(q_2,h) \, ,                  \eqno(15b)
$$
in analogy to $(11b)$. It describes the covariant extension of the
theory (2). Transition to the flat space is achieved by $h^\mp \to\pm 1$.

\subsection{4. Gauge equivalence between the WZNW system  
               and induced gravity} 

As a final step, we shall now construct a more general gauge
invariant extension of the WZNW system (2), and show that it reduces to
the induced gravity action (1) after a suitable gauge fixing. 

We first note that the currents $J_{(\mp)a}$ and $^*J_{(\mp)a}$,
defined by 
$$
^*J_{(\mp)\pm}\equiv J_{(\mp)\mp}\qquad 
                   ^*J_{(\mp)0}\equiv -J_{(\mp)0}\, ,  
$$
satisfy the same KM algebras. Now, we can use two sets of the KM
currents, corresponding to the L and R sectors of the WZNW theory
(2), to define new quantities
$$ 
I_{(\mp)a}=J^{(1)}_{(\mp)a} +^*J^{(2)}_{(\mp)a} \, .            \eqno(16a)
$$
The Poisson bracket algebra between $I_{(\mp)a}$ and $T_\mp$ has the
form  
$$\eqalign{
&\{T_\mp(\s_1),T_\mp(\s_2)\}=-[T_\mp(\s_1)+T_\mp(\s_2)]\d' \, ,\cr
&\{T_\mp(\s_1),I_{(\mp)a}(\s_2)\}=-I_{(\mp)a}(\s_1)\d'\, , \cr
&\{I_{(\mp)a}(\s_1),I_{(\mp)b}(\s_2)\}=f_{ab}{^c}I_{(\mp)c}(\s_2)\d \, , }
                                                                 \eqno(16b)
$$
representing two independent (left and right) semi--direct
products of the Virasoro and $SL(2,R)$ algebras. We see that the
currents $I_{(\mp)a}$ satisfy two $SL(2,R)$ algebras, since the
central charges of $J^{(1)}$ and $^*J^{(2)}$ {\it mutually
cancell\/}. Therefore, the set $(T_\mp, I_{(\mp)a})$ can 
be taken as a set of first class constraints, needed in the
construction of the canonical action.  In the theory (2) defined by a
difference of two WZNW actions, one is able to gauge the full
$SL(2,R)_{(-)}\times SL(2,R)_{(+)}$  symmetry, whereas for the simple
WZNW model for $SL(2,R)$ this is not possible.

We display here the complete set of constraints, multipliers and 
gauge parameters:
$$  
\vbox{\rm \halign{ # && $\,\,$ \hfil # \hfil  \cr 
$G_m=$      & $T_\mp$     & $I_{(\mp)a}$      \cr
$\,u^m=$    & $h^\mp$     & $A^{(\mp)a}$      \cr
$\,\ve^m=$  & $\ve^\mp$   & $\eta^{(\mp)a}$   \cr } } 
$$
In the canonical theory defined by the full set of constraints, the
gauge transformations of the multipliers $h^\mp$ are the same as in
$(12b)$, while those of $A^{(\mp)a}$ are given as
$$
\d A^{(\mp)a}=\pa_0\eta^{(\mp)a} +h^\mp\pa_1\eta^{(\mp)a} 
  +f_{bc}{^a}A^{(\mp)c}\eta^{(\mp)b} -\ve^\mp A^{(\mp)a} \, . 
$$

For our purposes, it will be enough to consider the restriction of the
above theory, defined by the subset of first class constraints
$$
T_\mp\, ,\,\, I_{(-)+}\, ,\,\, I_{(-)0}\, ,
         \,\, I_{(+)-}\, ,\,\, I_{(+)0}\, ,
$$
representing a subalgebra of $(16b)$. The canonical action of the
restricted theory takes the form 
$$\eqalign{
\cL(q_i,\pi_i,h)=&\pi_{1\a}\dot q_1^\a+\pi_{2\a}\dot q_2^\a  
                                               -h^-T_- -h^+T_+\cr 
    & -A^{(-)+}I_{(-)+} -A^{(-)0}I_{(-)0}
      -A^{(+)-}I_{(+)-} -A^{(+)0}I_{(+)0} \, .}              \eqno(17)
$$
The transformation rules for the multipliers are easily obtained from
the general expressions, by imposing the restriction 
$\eta^{(\mp)\mp}=0$, $A^{(\mp)\mp}=0$.

It is clear that the action (17) represents a generalization of the
(L--R) theory (2). Indeed, if we fix the gauge so that
$A^{(\mp)\pm}=A^{(\mp)0}=0$, the above theory reduces to
(15). Now, we shall demonstrate that a different gauge choice
reduces the action (17) to the form (1), proving thereby {\it the gauge
equivalence\/} between the WZNW theory (2) and the induced gravity (1). 

Let us fix the gauge symmetry corresponding to the first class
constraints $I_{(\mp)\pm},I_{(\mp)0}$,  by choosing the following gauge
conditions:  
$$
\O_{(\mp)\pm}\equiv J^{(1)}_{(\mp)\pm}-\m_{(\mp)} =0\, ,\qquad
\O_{(\mp)0}\equiv J^{(2)}_{(\mp)0}-\l_{(\mp)} =0 \, .          \eqno(18)
$$ 
To impose these gauge conditions in the functional integral, we
introduce a set of ghost fields $c^m=(e^\mp,c^{(\mp)a})$, 
antighosts $\bar c^{(\mp)a}$, and multipliers $b^{(\mp)a}$.
After introducing the gauge fermion $\Psi$ in the usual way, 
$$
\Psi= \bar c^{(-)+}\O_{(-)+} +\bar c^{(-)0}\O_{(-)0}
     +\bar c^{(+)-}\O_{(+)-} +\bar c^{(+)0}\O_{(+)0} \, ,      
$$
the gauge fixing and the Fadeev--Popov parts in the quantum action are
determined by 
$$\eqalign{
s\Psi=&\cL_{GF}+\cL_{FP}\, ,\cr 
\cL_{GF}=&b^{(-)+}\O_{(-)+} +b^{(-)0}\O_{(-)0} 
         +b^{(+)-}\O_{(+)-} +b^{(+)0}\O_{(+)0} \, , \cr
\cL_{FP}=&-\bar c^{(-)+}\bigl[s\O_{(-)+}\bigr]
          -\bar c^{(-)0}\bigl[s\O_{(-)0}\bigr]
          -\bar c^{(+)-}\bigl[s\O_{(+)-}\bigr]
          -\bar c^{(+)0}\bigl[s\O_{(+)0}\bigr] \, ,}          
$$
where $sX$ denotes the BRST transformation of $X$, obtained by
replacing gauge parameters in $\d X$ with ghosts.
Explicite form of the Faddeev--Popov term is obtained with the help of 
$$\eqalign{
&s\O_{(\mp)\pm} =-\bigl( e^\mp J^{(1)}_{(\mp)\pm}\bigr)' \mp 
                                       c^{(\mp)0}J^{(1)}_{(\mp)\pm}\, ,\cr
&s\O_{(\mp)0} =-\bigl( e^\mp J^{(2)}_{(\mp)0}\bigr)' 
  \mp c^{(\mp)\pm}J^{(2)}_{(\mp)\mp} \pm 2\k \bigl(c^{(\mp)0}\bigr)'\, . }
$$

The integration over the multipliers $A^{(\mp)}$ and $b^{(\mp)}$
yields 
$$
\cL(\vphi_i,\pi_{i\vphi},h)
  =\pi_{1\a}\dot q_1^\a+\pi_{2\a}\dot q_2^\a
   -h^-T_- -h^+T_+ +\cL_{FP}\ver{4}_{I=\O=0} \, .         \eqno(19)
$$
To evaluate this Lagrangian it is convenient to rewrite the first class
constraints $I_{(\mp)\pm}=0$, $I_{(\mp)0}=0$ and the related gauge
conditions in the form 
$$
J^{(1)}_{(\mp)\pm}=\m_{(\mp)} = -J^{(2)}_{(\mp)\mp} \, ,\qquad
  J^{(1)}_{(\mp)0}=\l_{(\mp)} = J^{(2)}_{(\mp)0} \, .   
$$
From here we see that $\pi_{1x},\pi_{1y}$ and $\pi_{2x},\pi_{2y}$ are
constants, therefore the related $\pi\dot q$ terms in the action can be
ignored as total time derivatives. The above conditions also ensure
that the contribution of the Faddeev--Popov term is decoupled from the
rest, so that the integration over antighosts and ghosts can be
absorbed into the normalization of the functional integral. The
calculation of $T_\mp$ leads to 
$$
\k T_\mp =\bigl[\pm(K_{1\mp})^2 +2\k(K_{1\mp})'\bigr] +
       \bigl[\mp(K_{2\pm})^2 +2\k(K_{2\pm})'\bigr]
       \mp{\fr 1 4}\m^2\bigl( e^{\vphi_1} -e^{\vphi_2}\bigr)\, ,   \eqno(20)
$$
where $K_{i\mp}=(\pi_{i\vphi} \mp\k\vphi_i')/2$, and $\m^2=\m_{(-)}\m_{(+)}$. 

As before, we eliminate the remaining momenta by using their equations
of motion,  
$$
\pi_{1\vphi}={\k\over h^--h^+}\bigl[ 2\dot\vphi_1 
               +\vphi_1'(h^-+h^+)+2(h^-+h^+)'\bigr] \, ,
$$
and $\pi_{2\vphi}=-\pi_{1\vphi}\vert_{\vphi_1\to\vphi_2}$, and obtain
the effective Lagrangian 
$$\eqalign{
&\cL(\vphi_1,\vphi_2,h)=\L(\vphi_1,h)-\L(\vphi_2,h) \, ,\cr
&\L(\vphi,h)\equiv {\k\over h^--h^+}\Bigl\{ 
     (\dot\vphi +h^-\vphi')(\dot\vphi +h^+\vphi')+\cr 
&\hskip45pt 2\bigl[ (h^-)'(\dot\vphi+h^+\vphi') 
            +(h^+)'(\dot\vphi+h^-\vphi') \bigr] \Bigr\} 
            +{\m^2\over\k}(h^--h^+)e^{\vphi}  \, . }           \eqno(21a)
$$
Now, if we introduce new variables $\phi$ and $F$,
$$
\phi={\a\over 2}(\vphi_1 -\vphi_2)\, ,\qquad F={1\over 2}\vphi_2 \, ,
                                      \qquad \a\equiv 2\sqrt{\k}\, ,
$$ 
the Lagrangian $\cL(\vphi_1,\vphi_2,h)$ can be written as a sum of
three terms:  
$$\eqalign{
&\cL_1={1\over h^--h^+}(\dot\phi+h^-\phi')(\dot\phi+h^+\phi')\, ,\cr
&\cL_2=\a \bigl(\o_0\phi'-\o_1\dot\phi\bigr) \, ,\cr
&\cL_3={\m^2\over\a^2}(h^--h^+)e^{2F} \bigl(e^{2\phi/\a}-1\bigr)\, , }
                                                             \eqno(21b)
$$
where
$$\eqalign{
&\o_0={1\over h^--h^+}\bigl[ (h^-h^+)'+2h^-h^+F'+(h^-+h^+)\dot F\bigr]\, ,\cr
&\o_1=-{1\over h^--h^+}\bigl[ (h^-h^+)'+(h^-+h^+) F' +2\dot F \bigr]\, . }
$$
To recognize the geometrical meaning of the action $(21b)$ 
we now introduce, in addition to the metric density $\tilde g^{\a\b}$,
Eq.$(12a)$, the determinant of the metric: 
$$
\sqrt{-g}={\fr 1 2}(h^--h^+)e^{2F} \, .                       \eqno(22)
$$
The transformation rule for $\sqrt{-g}$ is of the correct form. 
By noting the identity $\,\sqrt{-g}\,R=2(\pa_0\o_1-\pa_1\o_0)$,
one finds, after a partial integration in $\cL_2$, that the final form
of the action coincides with the induced gravity action (1), 
with $M^2=2\m^2/\a^2$.

\subsection{5. Concluding remarks} 

The general method of constructing canonical gauge invariant actions
is used to prove that the $SL(2,R)$ invariant WZNW system of the
(L--R) type, Eq.(2), is gauge equivalent to the induced 2D gravity (1). 

We first obtained the covariant extension $(15)$ of the (L--R) WZNW
theory, working with the local form of the action, and using the
energy--momentum components $T_\mp$ as the generators of the
diffeomorfisms. Then, we  constructed a more general gauge invariant
action (17), based on the set of first class constraints $T_\mp$ and
$(I_{(-)+},I_{(-)0},I_{(+)-},I_{(+)0})$, where $I$'s represent a
subalgebra of $SL(2,R)_{(-)}\times SL(2,R)_{(+)}$. If the gauge
symmetry is fixed in the simplest way, $A^{(\mp)}=0$, $h^\mp=\pm 1$,
the theory (17) goes over into the original (L--R) WZNW action (2),
while a different gauge fixing of the same symmetry, given by
Eq.(18), leads to the induced gravity action (1).  The induced
gravity and the (L--R) WZNW system for $SL(2,R)$ are thus shown 
to be gauge equivalent.

\vfill\eject

\subsection{References:}

\item{[1]} A. M. Polyakov, Phys. Lett. {\bf B103} (1981) 207.
\item{[2]} A. M. Polyakov, Mod. Phys. Lett. {\bf A2} (1987) 893.
\item{[3]} V. G. Knizhnik, A. M. Polyakov and A. B. Zamolodchikov, 
   Mod. Phys. Lett. {\bf A3} (1988) 819.
\item{[4]} O. Alvarez, Nucl. Phys. {\bf B216} (1983) 125.
\item{[5]} D. Friedan, in {\it Recent Advances in Field Theory and
   Statistical Mechanics\/}, Les Hou\-ches, 1982, eds.
   J. B. Zuber and R. Stora (North Holland, 1984) p. 839.
\item{[6]} A. M. Polyakov, Int. J. Mod. Phys. {\bf A5} (1990) 833.
\item{[7]} M. Bershadsky and H. Ooguri, Comm. Math. Phys. 
   {\bf 126} (1989) 49.
\item{[8]}  A. Alekseev and S. Shatashvili, Nucl. Phys. {\bf B323}
   (1989) 719.
\item{[9]} P. Forg\'acs, A. Wipf, J. Balog, L. Feh\'er and 
   L. O'Raifeartaigh, Phys. Lett. {\bf B227} (1989) 214.
\item{[10]} M. Henneaux and C. Teitelboim, {\it Quantization of Gauge
   Systems\/} (Princeton University Press, 1992). 
\item{[11]} B. Sazdovi\'c, Phys. Lett. {\bf B352} (1995) 64.
\item{[12]} A. Mikovi\'c and B. Sazdovi\'c, Mod. Phys. Lett. {\bf A10}
   (1995) 1041.

\bye